\documentclass[a4paper]{ESASPStyle}
\usepackage{epsfig}

\def\fig{.}

\def\note #1]{{\bf #1]}}

\def\dd{{\rm d}}
\def\Msun{M_\odot}
\def\gamp{\gamma,_p}
\def\gamrho{\gamma,_\rho}
\def\gamY{\gamma,_Y}

\newcommand{\pmbol}[1]{\setbox0=\hbox{#1}%
 \kern-0.018em\copy0\kern-\wd0
 \kern0.036em\copy0\kern-\wd0
 \kern-0.018em\raise0.028em\box0} 

\newcommand{\bnu}{\pmbol{$\nu$}}  
\newcommand{\bX}{\pmbol{$X$}}  

\begin{document}

\title{On Inverting Asteroseismic Data}

\author{Michael J. Thompson\inst{1}
\and J{\o}rgen Christensen-Dalsgaard\inst{2}}
\institute{ Space \& Atmospheric Physics, Blackett Laboratory, 
Imperial College, London, U.K
\and Teoretisk Astrofysik Center, Danmarks Grundforskningsfond, and \\
Institut for Fysik og Astronomi, Aarhus Universitet, {\AA}rhus, Denmark} 

\maketitle 

\begin{abstract}

Some issues of inverting asteroseismic frequency
data are discussed, including the use of model calibration and 
linearized inversion. An illustrative inversion of artificial 
data for solar-type stars, using least-squares fitting of a small
set of basis functions, is presented. A few details of kernel
construction are also given.

\end{abstract}

\section{Introduction}
\label{sec:intro}

The accumulated experience in helioseismology of inverting mode
frequency data provides a good starting point for asteroseismic inversion.
In some rather superficial ways the circumstances of the two may seem
very different:\break
helioseismologists can use  
many more mode frequencies than will ever be possible for a more
distant star, so that the resolution that can be achieved in helioseismic
inversion is beyond the grasp of asteroseismology 
(an exception may be 
situations such as where modes are trapped in a narrow range of depth 
within the star).
Another difference is that
global parameters of the Sun such as
its mass, radius and age are much better known than they are for other
stars; hence the structure of the Sun is constrained {\it a priori}
much more 
accurately than it is for a distant star, even if the input physics had been
known precisely, which of course is not the case for the Sun or for other stars.

In more fundamental ways, though, helioseismic and asteroseismic inversion
are much more similar than they are different.
Helioseismology has not
always been blessed with such a wealth of data, and helioseismologists in 
the early days of their subject learned the value of model calibration
and asymptotic description (in particular of low-degree modes) for making
inferences about the Sun. They also learned some of the
dangers and limitations of drawing inferences from real data. 
The modal properties in many asteroseismic targets will be similar to 
the low-degree solar modes. Moreover, the principles of 
inversion are the same in both fields: obtaining localized information 
about the unseen stellar interior; assessing resolution; taking account
of the effects of data errors; assessing what information the data really
contain about the object of study; judiciously adding additional 
constraints or assumptions in making inferences from the data.

This paper touches on a few of these points with regard to inverting
asteroseismic data, including the usefulness of optimally localized
averaging (OLA) kernels and model calibration using large and small
frequency separations for solar-type oscillation spectra. Much more
detail on those two topics can be found in the papers by 
Basu et al. (2001) and Monteiro et al. (2001), both in these proceedings. 

The inversion problem predicates that one is able to perform first
the forward problem. In the present
context, that means the computation of the oscillation data
(the mode frequencies) from the assumed structure of the star.
Solving the forward problem always involves approximations or 
simplifying assumptions, because we cannot model the full complexities of
a real star. The inverse
procedure, inferring the structure (or dynamics) of the star from the
observables, is ill-posed because, given one solution, there will almost
invariably formally be an infinite number of solutions that fit the 
data equally well. The art of inverse theory is in no small part concerned
with how to select from that infinity of possibilities. Actually, in
various asteroseismic applications to date, e.g. to $\eta$ Boo and 
certain $\delta$ Scuti stars,
the problem is rather that one has so far
been unable to find even a single solution that fits the data
(see Christensen-Dalsgaard, Bedding \& Kjeldsen 1995, 
Pamyatnykh et al.\ 1998).  This may 
indicate that the some of the approximations made in the forward 
problem are inappropriate; it could also indicate that 
the errors in the data have been assessed incorrectly.
In the present work, unless otherwise stated, we assume both that the forward 
problem can be solved correctly and that the statistical properties of the 
data errors are correctly known. 

%

\section{Model calibration}
\label{sec:calibration}

A relatively straightforward approach to the inverse problem is 
model calibration. At its conceptually simplest, this entails computing
the observables for a set of models, possibly a sequence in which one
or more physical parameters vary through a range of values, and choosing
from among that set the one that best fit the data. `Best' here is often taken
to mean that model which 
minimizes the chi-squared value of a least-squares fit to the data.
Given today's fast computers, searches through large sets of models are
feasible, either blindly or by using some search algorithm such
as genetic algorithm or Monte Carlo. 
Approaches similar to this have been undertaken 
for white dwarf pulsators 
(Metcalfe, Nather \& Winget 2000; Metcalfe 2001)
$\delta$ Scuti stars (Pamyatnykh et al. 1998), and 
pulsating sdB stars (Charpinet 2001). Depending on the way in which the
method is applied, the resulting model may be constrained to be a member of the
discrete set of calibration models, or could lie ``between'' them if 
interpolation in the set of models is permitted. 

Model calibration is powerful and dangerous. It is powerful because it allows
one to incorporate prejudice into the search for a solution and,
suitably formulated, it can always find a best fit. 
It is dangerous for the same reaons: perhaps unwittingly on the part of the 
practitioner, it builds prejudice into
the space of solutions that is considered. Also, even a satisfactory fit 
to the data does not mean that the solution model is necessarily like the
real star: one can make a one-parameter model calibration to a single
datum, but it is unclear what aspects of the resulting stellar
model the datum is actually able to constrain.
Examples of prejudice that model calibration may incorporate, for good or
ill, are the choice of physics used to construct the models, and 
possible assumptions about the smoothness of the structure of the 
star. 
An insidious problem is that, if the approximations in the forward 
modelling introduce errors into the computed model observables, 
this can introduce a systematic error into the result of the model calibration.
The Sun and solar-type stars provide a good example. Here the near-surface
structure and the mode physics in that region are poorly modelled at present,
and the simple approximations made introduce a systematic shift in the 
computed frequencies: low-frequency p modes have their upper turning point
relatively deep in the star and are almost unaffected by the treatment of
the surface layers, whereas modes of higher frequency have turning 
points closer to the surface and the error in their frequencies
grows progressively bigger. In these circumstances, it is preferable not
to calibrate to the frequency data themselves but rather to data
combinations chosen to be relatively insensitive to the 
known deficiency in the forward modelling. In the near-surface
layers, the eigenfunctions of solar p modes of low or intermediate degree $l$
are essentially independent of $l$, so the error in the frequencies
introduced by incorrectly modelling this region is just a function of
frequency, scaled by the inverse of the mode inertia. This suggests that
one should calibrate frequencies of solar-type stars to data combinations
chosen to be insensitive to such an error.

\begin{figure}[ht]
  \begin{center}
    \epsfig{file=\fig/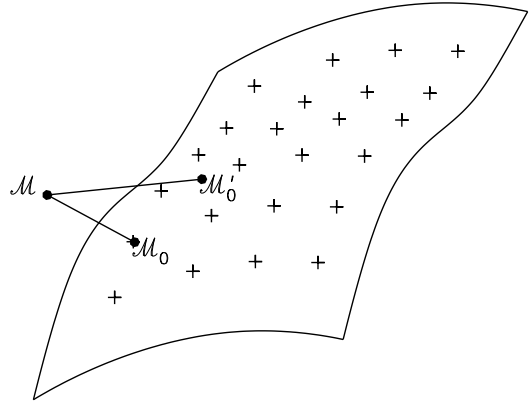, width=11cm}
  \end{center}
\caption{ Model calibration selects a model ${\cal M}_0$ in the 
space of models allowed by the calibration procedure: 
subsequent inversion using ${\cal M}_0$ as
reference determines a (hopefully improved) model $\cal M$, not
necessarily in the original space. If
some other calibration selected a different model ${\cal M}_0^\prime$,
one hopes that the inversion procedure would bring one to the same
or similar final model ${\cal M}$.
\label{fig:modelspace}}
\end{figure}

Asymptotic analysis suggests other data combinations which can be used for
model calibration and which are more discriminating than the raw 
frequency data. Monteiro et al. (2001) consider in some detail the use
of the so-called large and small separations:
\begin{equation}
\Delta_{nl} = \nu_{n+1\,l}-\nu_{nl}\;,\qquad
\delta_{nl} = \nu_{nl}-\nu_{n-1\,l+2}
\end{equation}
respectively. As Monteiro et al. demonstrate, the results of such a calibration,
even for such global characteristics as the mass and age of the star,
depend on the other physics assumed. 

\section{Linearized inversion}
\label{sec:ola}

Model calibration is just one approach to the inverse problem of inferring
the stellar structure, which we may indicate schematically as $\bX$,
from the frequency data $\bnu$. How else may the nonlinear
dependence ${\bnu} = {\bnu}({\bX})$ be inverted? 
A way which then allows application of a variety of techniques 
is to linearize about a reference 
model ${\cal M}_0$: inversion takes as input $\delta{\bnu}$, 
the difference between
the observed data and the corresponding values that model ${\cal M}_0$
predicts, and produces as output
$\delta\bX$, the estimated difference in structure between the 
star and ${\cal M}_0$.  
Since the structure of ${\cal M}_0$ is known, the structure of the star 
can then be reconstructed. Of course, that is a naive hope, because of
the inherent nonuniqueness discussed earlier, quite apart from
issues of data errors. But one may at least produce a refinement
on the initial model and indeed this new model can then be used as 
reference for a subsequent inversion as the next step of an iterative
approach. 
Depending on the technique adopted, it is helpful and may be essential 
to have a reasonable starting guess in the form of the reference model
${\cal M}_0$. Model calibration using the large and small separations 
is a reasonable way to find such a model for solar-type stars. 

Model calibration produces a solution ${\cal M}_0$
that is in the span (suitably
defined) of the calibration set of models, illustrated schematically
as a surface in Fig.~\ref{fig:modelspace}. 
Inversion of the kind just described may then
be used to proceed from ${\cal M}_0$ to a new model ${\cal M}$ which 
may be `close' to ${\cal M}_0$ but outside the span of the original
calibration models. 
One hopes that the final model depends only weakly on the initial model,
so that if by some other calibration (assuming different physics when
calibrating the large and small separations, for example) one produces
some other model ${\cal M}_0^\prime$, then the inversion step takes one
close once more to the same final model ${\cal M}$.

One can imagine making all structural
quantities in both the reference model and the target star 
dimensionless by taking out appropriate factors of the 
gravitational constant $G$ and the stars' masses $M$ and radii $R$. 
Then the frequency differences $\delta\omega_{nl}$ between the same mode
of order $n$ and degree $l$ in the two stars
can be related to the differences in dimensionless structural 
quantities by an equation such as
\begin{eqnarray}
\label{eq:diffeq}
\displaystyle{
{\delta\omega_{nl}\over\omega_{nl}}\ =\ }&
\displaystyle{
\delta\left(\ln \left({
GM\over R^3}\right)^\frac{1}{2}\right)} \nonumber 
\displaystyle{
\;+\;\int_0^1 K_{u,Y}^{(nl)}(x){\delta u\over u}(x)\,\dd x} \nonumber \\
&
\displaystyle{
\;+\;\int_0^1 K_{Y,u}^{(nl)}(x)\delta Y\,\dd x 
\;+\;{F(\omega_{nl})\over E_{nl}}}
\end{eqnarray}
where here the choice has been made to express the structural differences
in terms of $u$, the ratio of pressure $p$ to density $\rho$, and $Y$,
the helium abundance by mass: for discussion of this choice, see
Basu et al. (2001). Here, $\delta u$ is the 
difference in dimensionless $u$ between
the two stars, the differences being evaluated at fixed fractional
radius $x$ in the star. 
The first term on the right-hand side is a constant and just reflects the 
$(GM/R^3)^{1/2}$ homologous dependence of the frequencies.
The final term is some function of frequency, divided
by mode inertia $E_{nl}$, which 
absorbs uncertainties from the near-surface layers. For some details of
the construction of the kernel functions $K_{u,Y}^{(nl)}(x)$ and
$K_{u,Y}^{(nl)}(x)$, which are known functions derivable from
the reference model, see the Appendix. In an inversion, the left-hand 
side will be known; all terms on the right-hand side are to be inferred,
including the difference in $M/R^3$ between the reference model and 
target star.

\section{Structural and frequency differences}
\label{sec:models}

To motivate the rest of the paper we consider first the structural 
differences between a few stellar models. These will already indicate that
rather small uncertainties in global parameters of stars, e.g. mass or
age, can lead to large uncertainties in their structure. Of course the 
positive side of that is that the stellar structure is sensitive to 
those parameters and so, if the observable mode frequencies
are in turn sensitive to those aspects of the structure then we may have
some hope of using the observations to constrain e.g. the mass and age of
the star rather precisely.

\begin{figure}[ht]
\hskip -0.2cm 
    \epsfig{file=\fig/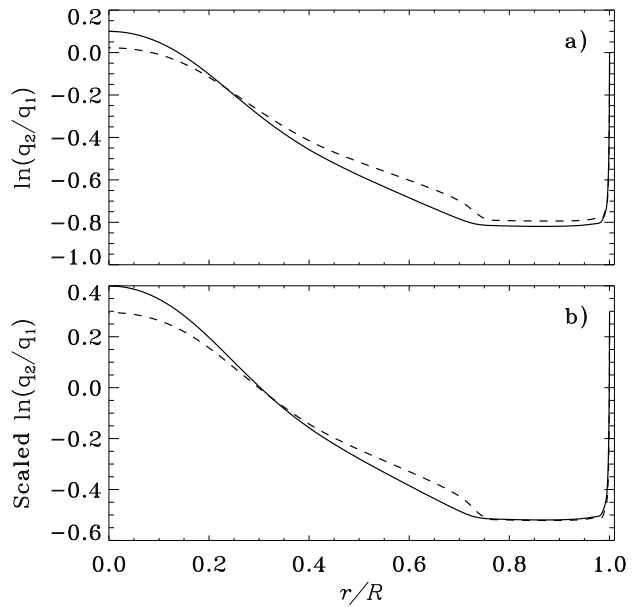, width=9cm}
\caption{ Relative differences in pressure $p$ (solid curve) and density $\rho$
(dashed curve), at fixed fractional radius, between ZAMS stars of mass
$1.0\Msun$ and $1.1\Msun$ (in the sense $1.1\Msun$ minus $1.0\Msun$). (a) The
quantities have not been homology-scaled before taking differences. (b)
Quantities have been homology-scaled before differencing.
\label{fig:prhodiffs}}
\end{figure}

Panel (a) of Figure~\ref{fig:prhodiffs} shows the relative differences in 
pressure $p$ and density $\rho$, at fixed fractional radius, between
two ZAMS stars, of masses $1.0\Msun$ and $1.1\Msun$.
We note that even for two stars with rather similar masses the
differences are large, of order unity. Panel (b) shows the corresponding
differences after the homology scaling has been taken out: this 
scaling is assumed taken out in the formulation presented in 
eq.~(\ref{eq:diffeq}). The effect is essentially to shift the two 
curves by a constant: although the differences are smaller, they are
still large. These changes arise from nonhomologous
differences in the surface layers which change the entropy of the 
convection zone. Indeed, writing $p=K\rho^{1+1/n}$ in the stars'
convective envelopes (in this context only, $n$ denotes polytropic index), 
and noting that sound speed there
is essentially determined by surface gravity, one finds that in that 
region
\begin{equation}
\delta\ln p \ \simeq\ \delta\ln\rho\ \simeq\ -n\,\delta\ln K\;.
\end{equation}

\begin{figure}[ht]
  \begin{center}
    \epsfig{file=\fig/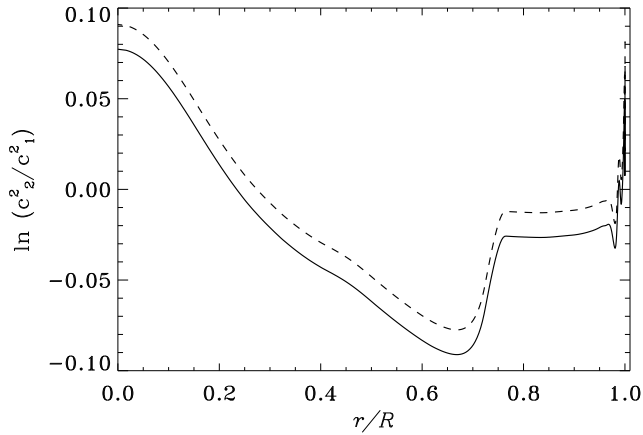, width=9cm}
  \end{center}
\caption{ Relative differences in sound speed squared ($c^2$) 
at fixed fractional radius, between ZAMS stars of mass
$1.0\Msun$ and $1.1\Msun$ (in the sense $1.1\Msun$ minus $1.0\Msun$): 
the two curves show
unscaled (solid) and homology-scaled (dashed) differences.
\label{fig:fig3}}
\end{figure}

Possibly of more direct relevance for inversion of eq.~(\ref{eq:diffeq})
is the difference in adiabatic sound speed $c$ ($c^2 = \gamma_1 u$, where
$\gamma_1$ is the first adiabatic exponent). As Fig.~\ref{fig:fig3}
shows, the relative differences in $c^2$ are smaller than the differences in 
pressure and density, but still quite substantial for stars that differ
in mass by as little 10 per cent. Homology scaling has little effect on 
the differences in this case. 
For a reasonable range of masses, these differences scale linearly with 
the mass difference, so for example the corresponding scaled differences between
$1.05\Msun$ and $1.0\Msun$ ZAMS stars are half those illustrated 
here, to a very good approximation.

\section{Inversion methods}
\label{sec:inversion}

A number of different linearized inversion methods have been developed in 
helioseismology, geophysics and diverse other areas of
inversion applications. Two flexible\break
 approaches are 
Optimally Localized Averages (OLA) and Regularized Least Squares
(see e.g. Christensen-Dalsgaard et al. 1990 for a description). The
OLA method explicitly constructs a linear combination of kernels
that is localized at some location in the star and is small elsewhere:
the corresponding linear combination of relative frequency differences 
is then a measure of the localized average of the structural differences.
The application of SOLA in the asteroseismology of solar-type stars is 
discussed and illustrated by Basu et al. (2001). We emphasize here that
OLA reveals the true extent to which the seismic data alone can
resolve the aspect of the stellar interior under study. Methods which 
on specific classes of problems appear to have superior performance to 
OLA in resolving aspects of the stellar interior are introducing 
nonseismic information or assumptions in addition to the frequency data.

The form of least-squares method used most extensively
in helioseismology
is regularized least-squares: the idea is to represent
the solution with a set of basis functions 
more finely than can be resolved by the data and with ideally
no bias about the form of the solution built into the basis; but then to 
minimize the sum of chi-squared fit to the data and a penalty term which
is large if the solution has undesirable characteristics. The most
used penalty function is the integral over the star of the 
squared second derivative of the function under study with respect to radius.

A different way of regularizing the least-squares solution, without introducing
a penalty term, is to choose a drastically smaller set of basis functions. 
This alternative, which we do not claim is intrinsically superior, may 
give apparently better results from few data if the basis functions are 
chosen with appropriate intuition or good fortune. The reason is that
one can introduce a huge amount of prejudice into the solution by 
forcing it to have a form determined by the basis functions. Such a 
basis could, for example, force the buoyancy frequency to be zero in a 
convective core and permit a single discontinuity at the core boundary but 
not elsewhere: such assumptions may be reasonable, but it should be 
realized that they are additional to the seismic data. If the star actually 
had a second discontinuity, or a more gradual variation at the core
boundary, such an inversion would not generally reveal those features.
A helioseismic example is in finding the location of the base of the 
convective envelope, where remarkable precision ($0.001R_\odot$ taking into
account uncertainties in abundance profiles, $0.0002R_\odot$ if the
only uncertainty 
comes from data noise) has been claimed (Basu \& Antia 1997): this
is credible only insofar as the base of the convection zone has precisely 
the form assumed in the inversion, because the true resolution at that
location is much coarser than that. The true resolution (e.g. vertically)
is essentially limited by the reciprocal of the largest vertical wavenumber of 
the eigenfunctions corresponding to the available data (Thompson 1993).

\section{An experiment with specific basis functions}
\label{sec:results}

As a simple illustration of the apparent ability of least-squares inversion
with a limited basis to infer structure even where the mode set has 
little resolving power, we take a basis of five functions chosen for 
algebraic convenience. They are approximately (but only approximately)
able to represent the difference in $u$ between ZAMS models of
solar-like stars. The basis
represents a function that is quadratic in $x\equiv r/R$
for $x<0.1$ with zero derivative at $x=0$, piecewise constant between
radii $x=0.1$, $0.3$ and $x_1$, quadratic between $x_1$ and $x_2$ and zero
for $x>x_2$. The values of $x_1$ and $x_2$, which are initially set to 
$0.65$ and $0.75$ respectively, are adjusted by hand to find a minimum
of the chi-squared fit.

\begin{figure}[ht]
  \begin{center}
    \epsfig{file=\fig/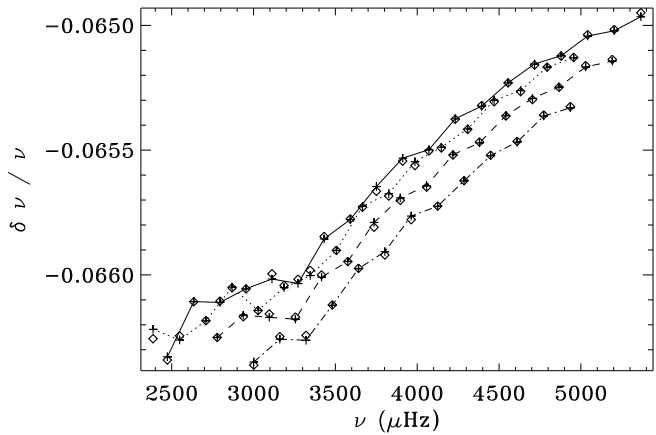, width=9cm}
  \end{center}
\caption{ Relative frequency differences ($\nu = \omega/2\pi$ is cyclic
frequency) between ZAMS stars of masses $1.0\Msun$ and $1.05\Msun$ (in 
the sense $1.05\Msun$ minus $1.0\Msun$). Lines join modes of the same
degree: $l=0$ (continuous line), $l=1$ (dotted), $l=2$ (dashed) and 
$l=3$ (dot-dashed). Crosses indicate the actual values of the 
frequency differences, diamonds indicate the values produced by a 
least-squares fit of expression (\ref{eq:diffeq}) using the basis 
functions discussed in Section~6.
.
\label{fig:fig4}}
\end{figure}

\begin{figure}[ht]
  \begin{center}
    \epsfig{file=\fig/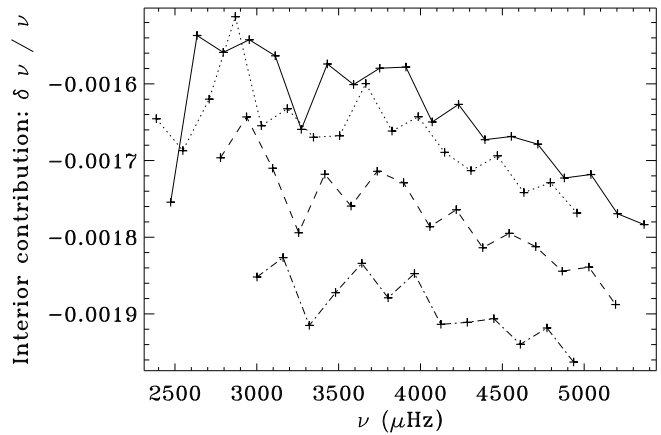, width=9cm}
  \end{center}
\caption{ Residual frequency differences after the fitted constant term
$\delta\ln(G M / R^3)^\frac{1}{2}$ and a two-term fit for
the surface-like contribution
$F(\omega)/E_{nl}$ have been removed. Line styles have the
same meaning as in Fig.~\ref{fig:fig4}.
\label{fig:fig5}}
\end{figure}

In the first of two illustrative applications, we consider the inversion
of frequency data (65 modes: 
$l=0$, $n=14-32$; $l=1$, $n=13-29$, $l=2$, $n=15-30$,
$l=3$, $n=16-28$) from a $1.05\Msun$ ZAMS star. In a real
application we would first calibrate the
star by computing large and small separations from the data
and using the approach of
Monteiro et al. (2001) to arrive at a reference model; in fact we 
simply took a reference model which was a $1.0\Msun$ ZAMS star.
Before discussing the results of the inversion, we first consider the 
data, i.e. the relative frequency differences, shown in 
Fig.~\ref{fig:fig4}. (For clarity, no noise has been added to the
data, though it was added for the inversions.) The dominant trends are
that the values are negative, around $-0.066$, because of the difference
in $M/R^3$ between the two stars (in fact, 
$\delta\ln((GM/R^3)^\frac{1}{2}) = -0.0668$ for these two stellar models);
and there is a roughly linear trend
with frequency, in this frequency range,
coming from near-surface differences. These two 
contributions can be estimated and removed
by fitting an expression of the form (\ref{eq:diffeq}): for illustration, 
we show in Fig.~\ref{fig:fig5} what would remain:
this is the signal from the interior, which contains the information 
that the inversion will use to infer conditions inside the star. The impression
now is that the data contain a signal which has some oscillatory component
(from the rather abrupt change in the structural 
differences at the base of the convective envelope) but is otherwise only 
weakly 
a function of frequency and increases with decreasing $l$, indicating that
the more deeply penetrating (i.e. lower-$l$ modes) sense $\delta u / u$ 
increasing 
in the deep interior as one gets closer to the centre of the star. 

\begin{figure}[ht]
  \begin{center}
    \epsfig{file=\fig/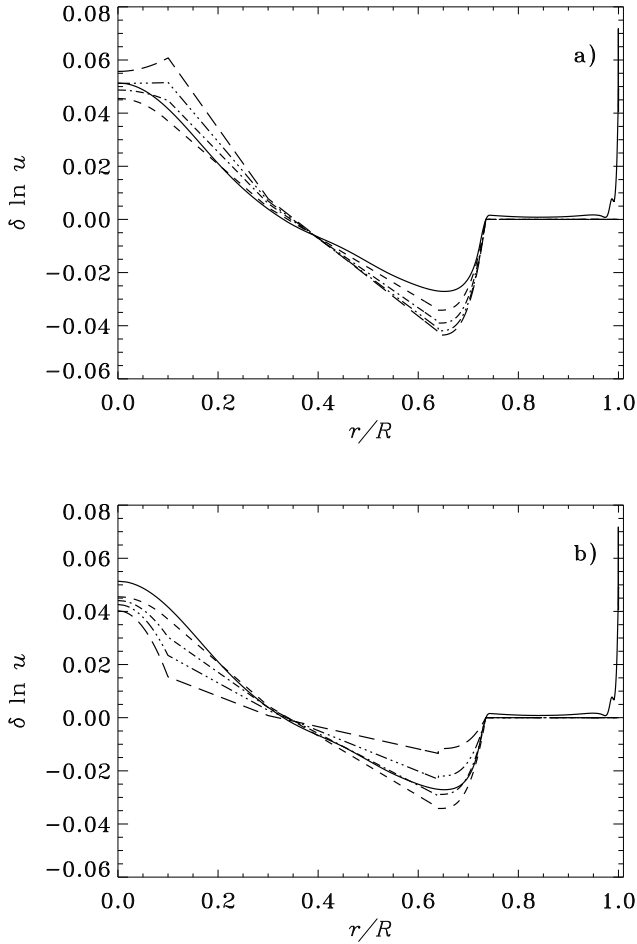, width=9cm}
  \end{center}
\caption{ Inversion results for the relative difference in homology-scaled
$u$ between a $1.05\Msun$ ZAMS star and a $1.0\Msun$ ZAMS star. The solid
curve shows the exact difference $\delta u/u$, and the short-dashed curve the
solution from a least-squares inversion of noise-free data. The other curves
show inversions of noisy data containing independent Gaussian errors 
with zero mean and uniform standard deviation $\sigma= 0.1\,\mu$Hz
(dot-dashed), $\sigma=0.2\,\mu$Hz (triple dot-dashed) and 
$\sigma=0.3\,\mu$Hz (long-dashed). The upper and lower panel show two 
different realizations of the noise; within each panel the noise from 
one curve to another has just been scaled by a multiplicative factor.
\label{fig:fig6}}
\end{figure}

These features are indeed revealed by the inversion (Fig.~\ref{fig:fig6}),
which compares the exact $u$-differences with the least-squares
solutions for noise-free data and for data with Gaussian noise with
zero mean and uniform standard deviations $\sigma \;=\;0.1\,\mu$Hz, 
$0.2\,\mu$Hz, and 
$0.3\,\mu$Hz. The noise realization in the two panels is different, but
within each panel the noise differs from case to case only by a multiplicative
scaling. It can be seen that for low noise levels this very small basis
enables the differences to be recovered rather well, including the structure
beneath the convection zone and the variation of $\delta u$ in the core. 
For larger noise levels the artificiality of the basis functions becomes 
more apparent; also the solution for the higher noise levels is rather 
different for the two noise realizations, which gives some indication 
of the uncertainty even in this highly constrained solution. 

\begin{figure}[ht]
  \begin{center}
    \epsfig{file=\fig/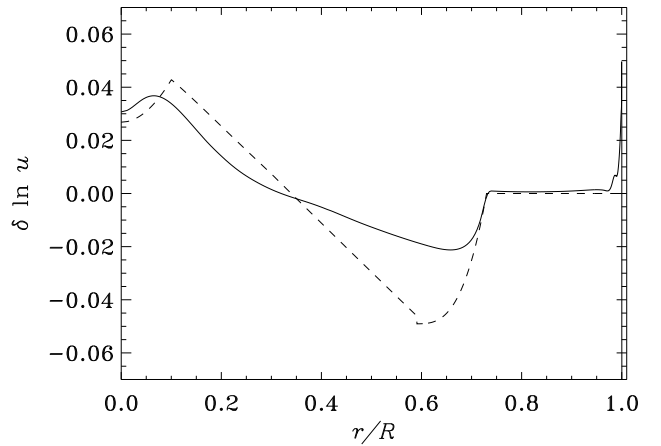, width=9cm}
  \end{center}
\caption{ Inversion results for the relative difference in homology-scaled
$u$ between model $S_1$ of Monteiro et al. (2001) and a present-day
standard solar model (in the sense $S_1$ minus solar model). The solid
curve shows the exact differences, the dashed curve shows the solution of
the least-squares fit to the data (which contained Gaussian noise with uniform
standard deviation $0.1\,\mu$Hz).
\label{fig:fig7}}
\end{figure}

A second example is an application to the same mode set but with data from
model $S_1$  of Monteiro et al. (2001) and Gaussian data errors with 
$\sigma=0.1\,\mu$Hz. This star is slightly more massive and
more evolved than the Sun. Again, we omitted the calibration step and
inverted relative to a reference model of the present-age Sun. The
results are shown in Fig.~\ref{fig:fig7}. Again the qualitative behaviour
of the differences is recovered reasonably, including the
downturn in the core. The discrepancies are perhaps partly attributable
to the fact that the basis functions are not so well suited for 
representing this case as the ZAMS case. 

\section{Discussion}
\label{sec:discussion}

It is a prejudice of some stellar astrophysicists (it was indeed expressed
a few times in Cordoba during the workshop) that helioseismic experience with
the Sun provides a poor example when it comes to asteroseismic inversion. 
But the similarities between the two applications are much more
significant than their differences. An extremely important aspect of 
inversion in any context is to assess what the data really tell you and
what information is being introduced by other assumptions or constraints.
OLA kernels (Basu et al. 2001) 
indicate what resolution can really be achieved without
additional assumptions.
But other approaches can advance our knowledge by allowing
introduction of reasonable prejudices: e.g., looking for signatures
of sharp features, or introducing specific basis functions. The best
apparent results are likely to be achieved if those functions are
physically motivated, because the solution will accord with 
our physical intuition (prejudice). This may of course be dangerous.
Our example of a highly constrained least-squares inversion illustrates
that qualitatively reasonable results can be obtained throughout a 
star by introducing assumptions about the form of the solution, even
in regions 
where in fact the mode set provides little or no localized information
(cf. Basu et al. 2001). One may have other grounds on which to believe that
such a solution is plausible, but on the basis of the data alone it
should be viewed sceptically.

Model calibration is a useful tool in its own right and for obtaining
possible starting models for linearized asteroseismic inversions. 
Carefully used, model calibration
allows one to build in some prejudice; and the combination of 
calibration and inversion extends the space of solutions that one explores.

Calibrating on large and small separations 
can be effective (Monteiro et al. 2001), always assuming that
one has not neglected some important aspect of the physics: in that
regard, the effects of rotation and magnetic fields need to be 
borne in mind
(see Dziembowski \& Goupil 1998), in solar-type stars as
in many other pulsating stars.

Finally we note that mode identification may be a problem, even in 
solar-type stars. The asymptotic pattern of high-order p-mode
frequencies may allow $l$ to be determined, but there may be some
uncertainty in $n$. Such an uncertainty can be allowed for in the 
inversion by adding to the right-hand side an extra term
\begin{equation} 
{G(\omega_{nl}) / \omega_{nl}}
\label{eq:neweq}
\end{equation}
where $G$ is some function of frequency. We have verified that, with our
ZAMS example, even restricting $G$ to be a constant function
removes quite satisfactorily the effect of a misidentification of $n$.
More generally, $G$ could judiciously be chosen to reflect the variation of the 
large separation with frequency.

\begin{acknowledgements}
This work was supported by the Danish National Research Foundation
through the establishment of the Theoretical Astrophysics Center,
and by the UK Particle Physics and Astronomy Research Council.
\end{acknowledgements}

\section*{Appendix}

The derivation of kernels relating the linearized differences in structure
to the differences in frequency (cf. eq.~\ref{eq:diffeq}) has been 
discussed by, e.g.,
Gough \& Thompson (1991), Gough (1993), 
and Kosovichev (1999). As all those authors show, 
the kernels for either of the pairs of variables $(c^2,\rho)$
or $(\gamma_1,\rho)$ are quite straightforward to derive from the 
equations of linear adiabatic oscillations together with the linearized
equation of hydrostatic support. Obtaining kernels for
various other pairs, including the pair $(u,Y)$ used in this paper, 
can be accomplished by first obtaining kernels for
one of the other two pairs and then using the following piece of manipulation.
It is sufficiently ubiquitous (occurring often when one wishes to transform
from kernels for a pair including $\rho$ to some other variable pair) that
we write it rather generally.

Let $\psi(r)$ be a solution of
\begin{equation}
\left({\psi^\prime\over r^2\rho}\right)^\prime\;+\;
{4\pi G\rho\psi\over r^2 p}\ =\ 
\left({F(r)\over r^2\rho}\right)^\prime\;,
\label{eq:a1}
\end{equation}
where prime denotes differentiation with respect to $r$ (or $x$ if
everything -- including $r$ -- is expressed in dimensionless variables),
for a given function $F(r)$ (so $\psi$ is a functional of
$F$),
with boundary conditions
$\psi(0)=0$ and $\psi(R)=0$. 
Then, provided $\delta m(0) = 0$ and $\delta m(R) = 0$, where $m(r)$ is
the mass interior to radius $r$, and letting
$\langle\dots\rangle$ denote integration from $r=0$ to $r=R$ ($x=0$ to $x=1$),
\begin{equation}
\left\langle
F(r){\delta\rho\over\rho}\right\rangle
\ \equiv\ 
\left\langle
-p\left(\psi/p\right)^\prime
{\delta u\over u}\right\rangle\;.
\end{equation}
Note that 
$\delta m(0) = 0$ holds in general; and 
in our present application, we scale all structural 
quantities by $G$, $M$ and $R$ to make them dimensionless, so 
$\delta m(R) = 0$ is forced to be true. These two conditions also
mean that $\left\langle 4\pi r^2\delta\rho\right\rangle$ is zero, so any 
multiple of $4\pi r^2\rho$ may be added to a kernel multiplying 
$\delta\rho/\rho$. Note that such an additional contribution to 
$F(r)$ makes no change to the right-hand side of eq.~(\ref{eq:a1}) and
hence the contribution to $\psi$ from such an addition is zero.

Obtaining kernels for $Y$ additionally requires an assumption about the
equation of state through $\gamma_1$, since the oscillations do not know 
directly about the chemical abundances. In the following we
write
\begin{equation}
\gamp \equiv \left({\partial\ln\gamma_1\over\partial\ln p}
\right)_{\rho,Y},\;
{\rm etc.},\ 
\gamY \equiv \left({\partial\ln\gamma_1\over\partial Y}
\right)_{\rho,p}\;.
\end{equation}
Then for convenience we record the following transformations:
\medskip
\begin{eqnarray}
(c^2,\rho)\;\rightarrow\;(u,\gamma_1) 
\nonumber\\
&K_{\gamma_1,u} \equiv K_{c^2,\rho}\nonumber\\
&K_{u,\gamma_1} \equiv K_{c^2,\rho}
-p\left(\displaystyle{{\psi\over p}}\right)^\prime
\end{eqnarray}
with $F\equiv K_{\rho,c^2}$;
\medskip
\begin{eqnarray}
(\gamma_1,\rho)\;\rightarrow\;(u,Y)
\nonumber\\
&K_{Y,u} \equiv \gamY K_{\gamma_1,\rho}\nonumber\\
&K_{u,Y} \equiv \gamp K_{\gamma_1,\rho}
-p\left(\displaystyle{{\psi\over p}}\right)^\prime
\end{eqnarray}
with $F\equiv (\gamp + \gamrho)K_{\gamma_1,\rho}
+K_{\rho,\gamma_1}$; and
\medskip
\begin{eqnarray}
(u,\gamma_1)\;\rightarrow\;(u,Y)
\nonumber\\
&K_{Y,u} \equiv \gamY K_{\gamma_1,u}\nonumber\\
&K_{u,Y} \equiv \gamp K_{\gamma_1,u}
+ K_{u,\gamma} -p\left(\displaystyle{{\psi\over p}}\right)^\prime\ 
\end{eqnarray}
with $F\equiv (\gamp + \gamrho)K_{\gamma_1,u}$.

\end{document}